# Attention Sensitive Web Browsing

**Joy Bose**
Samsung R&D Institute
Bangalore, India
joy.bose@samsung.com

**Amit Singhai**
Samsung R&D institute
Bangalore, India
a.singhai@samsung.com

**Anish Anil Patankar**
Samsung R&D institute
Bangalore, India
anish.p@samsung.com

**Ankit Kumar**
Samsung R&D institute
Bangalore, India
aki.ankit33@gmail.com

## ABSTRACT
With a number of cheap commercial dry EEG kits available today, it is possible to look at user attention driven scenarios for interaction with the web browser. Using EEG to determine the user's attention level is preferable to using methods such as gaze tracking or time spent on the webpage. In this paper we use the attention level in three different ways. First, as a control mechanism, to control user interface elements such as menus or buttons. Second, as a means to determine how much attention the user is paying to an area in the webpage, so the web developers can improve the webpage layout accordingly or insert ads in that section. Third, as a means for the web developer to control the user experience based on the level of attention paid by the user, thus creating attention sensitive websites. We present implementation details for each of these, using the NeuroSky MindWave sensor. We also explore issues in the system, and possibility of an EEG based web standard.

## ACM Classification Keywords
H.5.2. Information interfaces and presentation: User Interfaces - *Input devices and strategies, Standardization.*

## INTRODUCTION
While browsing the World Wide Web, the user may be more interested in certain sections of the webpage or in some webpages more than others. Knowledge of which sections are more interesting can help make websites attention sensitive and increase user engagement. When deciding which sections the user is interested in, current solutions generally focus on indicators like the dwell time i.e. time spent on the webpage and/or a particular section of the webpage. When deciding which sections the user is interested in, current solutions generally focus on indicators like the time spent in reading a section of the webpage or time spent looking at the section. However, the actual attention level of the user at the moment when they are reading the webpage is not necessarily correlated with these, since the user may not be paying attention even though they are looking at the page.

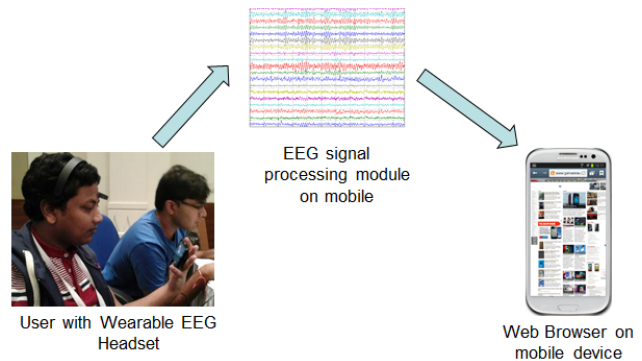

**Figure 1. High level architecture for attention sensitive web browsing**

Wearable and portable electroencephalography (EEG) kits can determine the momentary attention level of the user in real time, and transmit the data to a computing device such as a mobile phone using Bluetooth. Using such kits, it is possible to determine the sections of the web page where the user's attention level is high. Sending this data to the web content developers can provide valuable feedback to them, and help to design better websites.

In this paper, we explore the uses of knowing the attention level while browsing. We use the measured attention level in three ways: as a control mechanism for menus and other elements, as a browser level implementation for determining the area of interest to send feedback to the web developers, and as a means for web developers to control how the user experiences the website depending on their attention level.

One of the commercially available and cheap EEG kits is Neurosky's MindWave [1]. It consists of just one dry sensor, is lightweight and seems almost like a headset. Its algorithms process the raw EEG output from the forehead to produce a value for attention, meditation and blink. The attention meter algorithm [26] is used to get the attention in the NeuroSky headset. It has been used in a number of mainly gaming applications for user interactions based on attention. There have been comparatively fewer studies of using the headset in real life applications, perhaps due to a number of issues concerning accuracy and usability. In this paper we chose the Neurosky headset to measure attention, partly because of its comparatively cheap price, widespread usage, ready availability of SDKs, and ease of usage due to which it has a good chance of becoming a mass usage device.

Here we do not claim that our implementation is more accurate than other solutions; rather we agree that using research grade EEG kits such as B3000 would probably produce more accurate results. Our intention when writing this paper is simply to explore the ways in which attention can be used in web browsers, and make the argument that it is now feasible, and perhaps even desirable, to use attention as an extra input. Also, our targeted users are not just those who are paralyzed and have lost motor control, but also everyday users who might not be able to use their hands, such as when they are driving or waiting in a queue, and for whom attention may be an interesting way to input as a complement to other inputs.

One issue in the accuracy of EEG is the inevitable incidence of false positives. Our experimental setup is intended to be such that even with the occasional false positives generated by the EEG input, the overall user experience should not be dampened. Rebolledo-Mendez, G., I. Dunwell et al. [28] conducted a usability study on the Neurosky device and found a significant positive correlation between the attention levels measured by the device and the participants' self-reported attention values measured by a post-test questionnaire.

**RELATED WORK**
There are a number of wearable EEG sensors sold commercially and their popularity is increasing. Some such devices include NeuroSky's MindWave [1], Emotiv's EPOC and Insight [2], and Muse [3].

EEG sensors have been used to measure a user's momentary attention level in a number of different contexts [4-7]. Aside from gaming and learning applications, the EEG headset readings have also been used to change TV channels [8], control Google Glass [9], change a film's plot [10] and even for experimental art [11].

Previous work [24-25] as shown that the attention value can be extracted from the EEG sensors. The Neurosky EEG kit [26-27] uses eSense meters to measure the attention level as a scale of 1 to 100. The exact algorithm used by NeuroSky is not available, but the measured attention level has been previously used in a number of games and other applications.

A number of methods have been used to track a user's web page reading. Such methods include eye gaze tracking [12] and using JavaScript [13] to note the time spent on a section of the web page. Additionally, cursor movement, which is a good proxy for gaze tracking, and scrolling have also been used for the purpose [14-15]. Of these, eye gaze tracking and scrolling the web page in synchronization with the eye gaze has been reported to be mildly irritating to many users [16]. Other methods use the user's emotion [17-19] or perform data analytics on user behavior to determine the user's interest. As mentioned, such methods may not be accurate indicators of the user's momentary interest level, since the user might be staring at a webpage but be distracted thinking of something else.

For such reasons, our proposed EEG based solution is more accurate and a better choice to track the user's attention. Of course, it is quite possible that users may be concentrating on something different from the opened webpage. However, the attention level still makes for a decent approximation and can be combined with methods such as eye tracking to make sure the user is actually looking at the web page when they are paying attention to it.

A growing trend among web developers is that of responsive web design [20], where the webpage layouts etc. are modified dynamically with parameters such as the device display size or ambient light. This principle can be further improved by adding the attention level as a parameter for the responsive design.

**ATTENTION FOR CONTROL**
In the first of the three ways, we explore how the EEG signals can be used for controlling the functions of a web browser. This can be potentially useful for people who are immobile or paralyzed, or those who cannot temporarily use their hands e.g. while driving or waiting in a queue.

Here we mainly consider controls such as back and forward navigation buttons, menu navigation, and link selection. We intentionally do not include new URL entry, since pressing keys in a virtual keyboard with the help of the EEG attention level can be quite cumbersome. If totally hands free functionality is required, the URL entry can be performed through voice using a suitable Voice API.

We use the attention and blink levels obtained from the Neurosky EEG sensor using their provided SDK [21]. Other EEG kits such as Emotiv (in their Cognitiv suite) [2] have APIs where the system can detect the intention of the user to push or pull or move to the right or left. Some of those APIs might be better suited for a variety of control functions, but here we did not use them.

In our logic, we use the attention as a threshold for control functions, where the user navigates to the next item (similar to the focus movement with 'tab' key function in Windows) if the attention is higher than the threshold for a specific time, say 1 second. The attention level is received as a series of inputs, at 1Hz frequency, while the raw EEG data in the NeuroSky sensor is obtained at a 512 Hz frequency.

One problem with this setup is that the attention level fluctuates too quickly, and using the attention to achieve a regulated task such as controlling a web browser requires some amount of user familiarization. On the other hand, this method requires no training and only very little calibration, only to set the attention threshold for a particular user. This calibration can be performed by asking the user to perform a simple task or stay still for a few seconds while wearing the EEG headset, to set the baseline EEG attention level. Also, we noticed there was a slight delay between the users' changed attention and the time it was recorded by the EEG sensor.

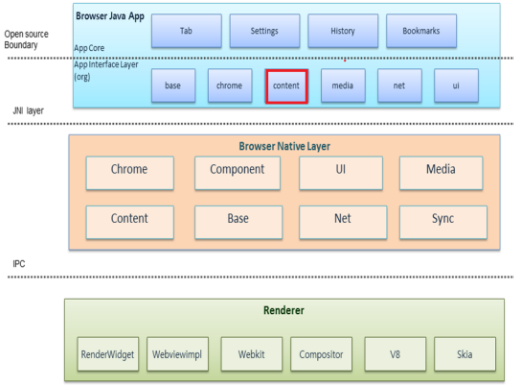

**Figure 2. Architecture of the mobile web browser showing changed components (in red)**

It is also true that the attention threshold for a user can vary with time of day and other contextual factors such as user location. For this we can either set up a regression model to learn and automatically modify the attention level when the same context is encountered, e.g. for a specific time of day, or perform some initial calibration.

**Implementation with the Neurosky Mindwave EEG kit**

As mentioned, the Neurosky Mindwave device measures raw EEG data at 512 Hz and outputs the attention value as an integer between 1 and 100, at 1 Hz, using custom algorithms to obtain e-Sense values of attention and meditation.

In our implementation, we use the EEG attention level for navigation across menus, buttons and other control elements in a webpage and the blink level (also detected by the EEG kit) to select items. Using the EEG input, the user navigates to items in the device home screen and selects a particular item to see the contents. We used the following protocol for webpage navigation: if the attention level remained over the threshold for 1 sec, we stay on the current item, else we move to the next item.

We start a new Activity in the Android system that listens for the EEG input from the NeuroSky sensor. The browser interface is implemented using a WebView in Android. We created two Activities: the Main Activity, corresponding to the browser home screen, and the Browser Activity. The Main Activity is visible to the user when the browser application is launched. Once the control item such as a button or menu item is selected by the user, the Browser Activity is launched and the URL associated with the selected item is loaded into the embedded web view. When the user does a deliberate double blink, they can return to the home screen or main activity. The double blink is also used to open a currently selected item in the web browser. We used deliberate double blinks to distinguish from casual blinks.

The NeuroSky sensor gives a numerical value between 0 and 100 to indicate the blink strength, and a similar value to indicate the attention level. A suitable threshold is used to detect a blink event and a high attention event.

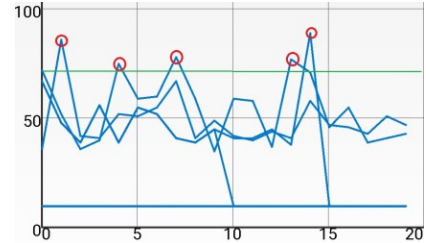

**Figure 3. Plot of the blink level of one user over a series of trials. Red circles denote deliberate double blinks.**

To determine the optimal threshold for blinks, we recorded the blink values for a series of trials on one user. The results are plotted in Fig. 3. We found that the difference between the current and previous blink values of 20 can distinguish between a casual and deliberate blink.

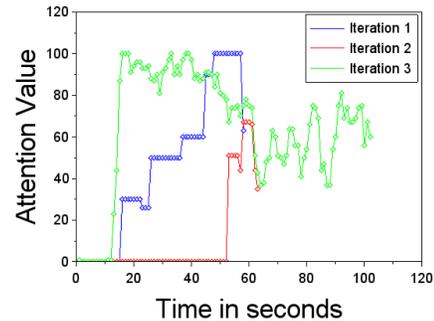

**Figure 4. Plot of a user's attention level while browsing through a Google news article, recorded over three iterations.**

To determine the attention threshold, we made a series of recordings of the attention level when a user was performing a news article reading activity on the web browser. Fig. 4 plots the results over 3 iterations. We found that for most users, the attention level consistently stayed over 30 when they were paying attention to the article.

**ATTENTION RESPONSIVE WEB BROWSER**

The attention level can also be used to inform the web content developers of which section or area of the webpage the user is more interested in, as a means for them to modify the layout of the webpage accordingly or to insert advertisements in these sections of the webpage. Here too, the application layer of the web browser is modified as shown in section 2.

We map the attention level to the section or block element (in HTML parlance) that the user is reading on the current webpage. The section is captured via the scroll position, determined using JavaScript. The script is executed in the context of the webpage to perform the tracking, along with a script to receive the attention inputs from the EEG sensor.

We calculate the average scroll percentage of the webpage using the following formula:

*Average scroll percentage = ((2\* Scroll offset + view port height) / content height) \* 100*

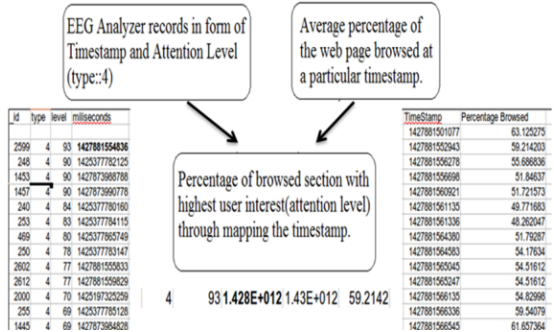

**Figure 5. Mapping the attention level and page scroll percentage readings, using timestamp as the key.**

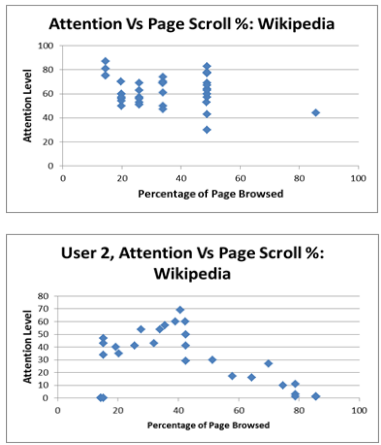

**Figure 6. Plot of the attention level with percentage of the Wikipedia homepage browsed for two users.**

The user's attention level is dumped to a file along with the percentage of the web page (scroll offset), along with the timestamp. In this way, we will have one file with mapping of percentage of page versus attention level of users. This is for pages where the user is reading some text. We later verified that the system correctly recorded the actual sections where the user was paying attention by asking the user if the reading reflected their interest.

Using this method, we plot the attention value together with the percentage of the current page scrolled, at a given time. Initially we stored the values of the EEG sensor and content script in two different tables and merged them offline using the time stamp as the key, as shown in Fig. 5.

Fig. 6 plots the attention level Vs scroll positions recorded for two users using the Wikipedia mobile homepage. We can see that the first user pays more attention towards the middle of the webpage and none at the end, while the second user's attention tapers more gradually.

**Sending aggregated data to the web content developers**

The user's interest level, recorded along with the webpage section, can be used as feedback to the web content developers in the following way:

The aggregated attention data can be sent to the web server. This can be used to improve the website design, or push ads or other relevant content on the sections of the webpage where most users are interested.

The aggregation of data is relevant where user privacy is more of an issue, so data can be sent in an encrypted form to the web server and stored in a way that does not expose the behavior of individual users.

**ATTENTION SENSITIVE WEB ARCHITECTURE**

This is the third way in which attention can be used while browsing. Here our goal is to enable web developers to make the websites attention sensitive, i.e. to be able to control how the user experiences the webpage depending on their level of attention. For example, the web developer may want a certain section of the webpage to be expanded if the user is paying attention, or different content to be generated and sent to the user's device. Our implementation involves the following steps, shown in fig. 7:

- Create class APIWrapperEEG class to receive and process events from the EEG device
- Annotate APIWrapperEEG class @JavascriptInterface
- Instantiate APIWrapperEEG on the application's Main Activity creation
- Find out page load finish event callback, Register APIWrapperEEG instance, with Content View on the callback using addJavaScriptInterface
- We also provide an EEG.js JavaScript file containing all helper methods which the developer can include and use in their web pages. These are basically wrapper functions and events for the APIWrapperEEG Object.

Regarding browser level changes, the browser engine code will remain the same, only there will be a glue code as part of the application layer to take inputs from the EEG sensor.

**Adapting existing web standards to EEG**

HTML5 has brought many modern standardized interfaces to web developers [20-21]. However, so far it does not support EEG. EEG based input mechanism is peculiar in that it has some characteristics of conventional input devices such as keyboard or mouse as well as those of continuous streaming input like microphone and camera.

One possible way to enable a specification for EEG is as follows: The fundamental block of the API is EEGStream interface. EEGStream consists of 0 or more EEGStreamTracks, each track representing each input signal from the EEG device available to the browser. Each EEGStreamTrack reads only attributes id and label, where id is a unique identifier representing the stream while label is a string human readable identifier for the stream. While creating an EEGStream object, an EEGConstraints object is also passed that specifies the frequency of the sampling and the callbacks associated with the same.

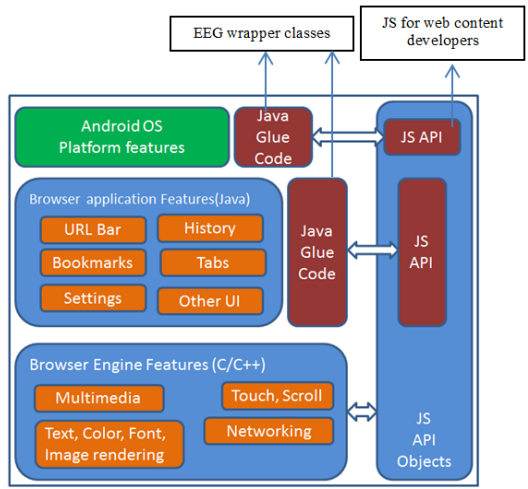

**Figure 7. Modifications in an Android system to enable attention sensitivity**

Higher level features such as blink detection can be implemented as a JavaScript library using the lower level API defined above. This would allow extensible and adaptable framework libraries to evolve for specific use cases while the lower level API remains generic. However, the EEG wearables area is still maturing and more implementation experience is needed before such a standard can be adopted.

## CONCLUSION AND FUTURE WORK

In this paper we have explored methods to enable attention sensitive web browsing. In future, we plan to extend the approach for other EEG sensed parameters such as the user's relaxation level and intention of movement.